\documentclass[journal]{IEEEtran}

\usepackage{amsfonts}
\usepackage{amsmath}
\usepackage{amssymb}
\usepackage{graphicx}
\usepackage{dsfont}
\usepackage{algorithm}
\usepackage{algorithmic}
\usepackage{multirow}
\usepackage{slashbox}
\usepackage{enumerate}
\usepackage{subfigure}
\usepackage{stfloats}

\graphicspath{{Figures/}}
\DeclareGraphicsExtensions{.eps}

\normalsize

\begin{document}
%
\title{Space-Time Polar Coded Modulation}
%
%
%

\author{Kai~Chen,~\IEEEmembership{Student Member,~IEEE,}
        Kai~Niu,~\IEEEmembership{Member,~IEEE,}
        and~Jiaru~Lin,~\IEEEmembership{Member,~IEEE}
\thanks{K.~Chen, K.~Niu and J.~Lin are with the Key Laboratory of Universal Wireless Communications, Ministry of Education, Beijing University of Posts and Telecommunications, Beijing 100876, China (e-mail: \{kaichen, niukai, jrlin\}@bupt.edu.cn)}%
}


\maketitle

\begin{abstract}

The polar codes are proven to be capacity-achieving and are shown to have equivalent or even better finite-length performance than the turbo/LDPC codes under some improved decoding algorithms over the additive white Gaussian noise (AWGN) channels.
Polar coding is based on the so-called channel polarization phenomenon induced by a transform over the underlying binary-input channel.
The channel polarization is found to be universal in many signal processing problems and has been applied to the coded modulation schemes.
In this paper, the channel polarization is further extended to the multiple antenna transmission following a multilevel coding principle.
The multiple-input multile-output (MIMO) channel under quadrature amplitude modulation (QAM) are transformed into a series of synthesized binary-input channels under a three-stage channel transform.
Based on this generalized channel polarization, the proposed space-time polar coded modulation (STPCM) scheme allows a joint optimization of the binary polar coding, modulation and MIMO transmission.
In addition, a practical solution of polar code construction over the fading channels is also provided, where the fading channels are approximated by an AWGN channel which shares the same capacity with the original.
The simulations over the MIMO channel with uncorrelated Rayleigh fast fading show that the proposed STPCM scheme can outperform the bit-interleaved turbo coded scheme in all the simulated cases, where the latter is adopted in many existing communication systems.

\end{abstract}

\begin{IEEEkeywords}
Polar codes, space-time coding, coded modulation, multilevel coding, joint optimization.
\end{IEEEkeywords}

%
\IEEEpeerreviewmaketitle

\section{Introduction}

\IEEEPARstart{P}{olar} codes are the first structured codes (as opposed to random codes) that provably achieve the symmetric capacity of binary-input memoryless channels (BMCs) \cite{Arikan}.
This capacity-achieving code family is based on a technique called channel polarization.
Given a BMC $W$, after performing the channel transform, which consists of the channel combining and splitting operations, over a set of independent copies of $W$, a second set of synthesized channels is obtained.
As the transformation size, i.e., the number of channel uses participated in the transform, goes to infinity, some of the resulting channels tend to be completely noised, and the others tend to be noise-free, where the fraction of the \mbox{noise-free} channels approaches the symmetric capacity of $W$.
By transmitting free bits over the noiseless channels and sending fixed bits over the others, polar coding with a very large code length $N$ can achieve the symmetric capacity under a successive cancellation (SC) decoder with both encoding and decoding complexity $O\left( N\log N \right)$.
To construct a polar code, the capacities (or equivalently, reliabilities) of the polarized channels can be estimated efficiently by calculating Bhattacharyya parameters for binary-input erasure channels (BECs) \cite{Arikan}.
But for channels other than BECs, computationally expensive solutions based on density evolution (DE) \cite{DE} and other modified methods are required to calculate the channel reliabilities \cite{construct}, \cite{construct_on}.

Although polar codes are asymptotically capacity achieving, the performance under the SC decoding is unsatisfying in the practical cases with finite-length blocks. Several improved SC decoding schemes have been proposed to improve the finite-length performance of polar codes.
The successive cancellation list (SCL) decoding \cite{SCL:Tal}, \cite{SCL} and successive cancellation stack (SCS) \cite{SCS} decoding algorithms are introduced to approach the performance of ML decoding with acceptable complexity.
By regarding the improved SC decoding algorithms as path search procedures on the code tree, the SCL and SCS decoding are the ``width-first'' and the ``best-first'' search, respectively.
To provide a flexible configuration under the constraint of both the time and space complexities, an decoding algorithm called the successive cancellation hybrid (SCH) is proposed by combining the principles of SCL and SCS \cite{imdec_TCOM}.
Moreover, under these improved SC decoding algorithms, polar codes are found to be capable of achieving the same or even better performance than turbo codes or low-density parity-check codes with the help of cyclic redundancy check (CRC) codes \cite{list_arxiv} \cite{acadec} \cite{CAdec}.
Therefore, polar codes are believed to be competitive candidates in future communication systems.

Shortly after the polar code was firstly put forward, the channel polarization phenomenon has been found to be universal in many other signal processing problems, such as multiple access communications \cite{mac1} \cite{mac2}, source coding \cite{source_coding1} \cite{source_coding2}, information secrecy \cite{security1} \cite{security2} and other settings.
To improve the spectral efficiency, a $2^m$-ary polar coded modulation scheme is provided in \cite{pcm}. By regarding the dependencies between the bits which are mapped to a single modulation symbol as a kind of channel transformation, the polar coded modulation (PCM) scheme is derived under the framework of multilevel coding \cite{multilevel_codng}.
It is shown in \cite{opcm} that this polar coded modulation scheme can outperform the turbo coded modulation scheme in the 3GPP WCDMA system \cite{WCDMA} by up to $1.5$dB with $64$-ary quadrature amplitude modulation (QAM) over additive white Gaussian noise (AWGN) channel.

In this paper, the channel polarization technique is extended to the multiple-input multiple-output (MIMO) transmission scenario.
Similar to the polar coded modulation scheme \cite{pcm}, the transmission over the MIMO channel is further combined into the channel transform. The MIMO transmission, modulation and the conventional binary channel polarization form a three-stage channel polarization procedure.
Based on this generalized channel polarization, a jointly optimized space-time polar coded modulation (STPCM) scheme is proposed.

The remainder of the paper is organized as follows.
\mbox{Section \ref{section_model}} introduces the system model concerned in this paper.
\mbox{Section \ref{section_polarization}} provides a three-stage channel transform which can be seen as a joint processing of the conventional binary channel polarization, modulation and MIMO transmission;
\mbox{Section \ref{section_stpcm}} describes the construction, encoding and decoding of the proposed STPCM scheme.
\mbox{Section \ref{section_simulation}} evaluates the performance of the proposed STPCM scheme under the Rayleigh fading channel through simulations.
Finally, \mbox{Section \ref{section_conclusion}} concludes the paper.

\section{Notations and System Model}
\label{section_model}

\subsection{Notation Conventions}

In this paper, the capital roman letters, e.g. $X$, $Y$, are used to denote random variables.
The lowercased letter $x$ denotes a realization of $X$.
${\Re}(x)$ and ${\Im}(x)$ are the real and image parts of a complex number $x$, respectively.
The modulus of $x$ is written as $\|x\|=\sqrt{\Re(x)^2+\Im(x)^2}$.
The calligraphic characters, such as $\mathcal{X}$ and $\mathcal{Y}$, are used to denote sets, and we use $|\mathcal{X}|$ to denote the number of elements in $\mathcal{X}$.
The Cartesian product of $\mathcal{X}$ and $\mathcal{Y}$ is written as $\mathcal{X} \times \mathcal{Y}$, and ${\mathcal{X}}^n$ stands for the $n$-th Cartesian power of $\mathcal{X}$.

We use notation $v_1^N$ to denote an $N$-dimensional column vector $\left(v_1, v_2, \cdots, v_N \right)$ and $v_i^j$ to denote a subvector $\left(v_i, v_{i+1},\cdots, v_{j-1}, v_j\right)$ of $v_1^N$, $1\leq i,j \leq N$.
When $i>j$, $v_i^j$ is a vector without elements, and this empty vector is denoted by $\phi$.
We write $v_{1,o}^N$ to denote the subvector of $v_1^N$ with odd indices ($v_k: 1 \leq k \leq N$; $k$ is odd).
Similarly, we write $v_{1,e}^N$ to denote the subvector of $v_1^N$ with even indices ($v_k: 1 \leq k \leq N$; $k$ is even).
For example, for $v_1^4$, $v_{2}^3=(v_2,v_3)$, $v_{1,o}^4=(v_1,v_3)$, and $v_{1,e}^4=(v_2,v_4)$.
Further, given an index set $\mathcal{A}$, let $v_{\mathcal{A}}$ denote the subvector of $v_1^N$, which consists of $v_i$s with $i \in \mathcal{A}$.

The matrices are denoted by bold letters, e.g., $\mathbf{X}$. The notations $\mathbf{X}^\prime$ and $\mathbf{X}^\dag$ stand for the transpose and conjugate transpose of $\mathbf{X}$, respectively.
The element in the $i$-th row and the $j$-th column of matrix $\mathbf{X}$ is written as $x_{i,j}$. The $j$-th column of matrix $\mathbf{X}$ is written as $\mathbf{X}_j$; the $i$-th row of $\mathbf{X}$ is written as $\mathbf{X}^\prime_i$, i.e., the $i$-th column of $\mathbf{X}^\prime$.
Furthermore, we write $\mathbf{F} \otimes \mathbf{G}$ to denote the Kronecker product of two matrices $\mathbf{F}$ and $\mathbf{G}$, and ${\mathbf{F}}^{\otimes n}$ to denote the $n$-th Kronecker power of $\mathbf{F}$.

Throughout this paper, $\log$ means ``logarithm to base 2'', and $\ln$ stands for the natural logarithm.

\subsection{System model}

\begin{figure}[!t]
  \centering
  \includegraphics[width=0.95\columnwidth]{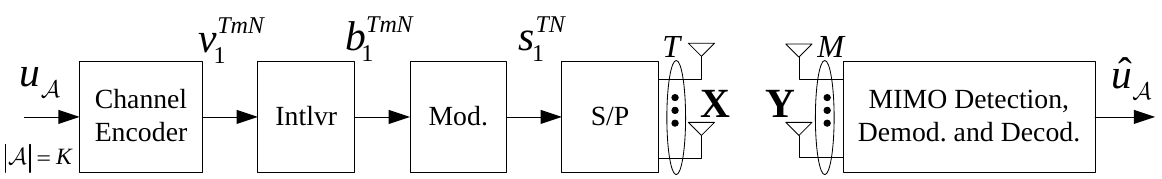}
  \caption{Block diagram of space-time coded modulation.}
  \label{fig_model}
\end{figure}

A block diagram of space-time coded modulation is depicted in \mbox{Fig. \ref{fig_model}}. The $K$ information bits are coded and modulated into a series of $2^m$-ary symbols, and then transmitted to the receiver through a MIMO system with \mbox{$T$-transmit} and \mbox{$M$-receive} antennas within $N$ time slots.

At the transmitter, a sequence $u_{\mathcal{A}}$ of $K$-length information bits, where $|\mathcal{A}|=K$, is fed into a binary channel encoder with code rate $R=\frac{K}{TmN}$.
The encoded sequence $v_1^{TmN}$ is interleaved into an other binary sequence $b_1^{TmN}$.
After a \mbox{$2^m$-ary} modulation, the $TmN$ bits are mapped into $TN$ complex symbols $s_1^{TN}$.
These symbols are then partitioned into $T$ streams with $N$ symbols in each stream and respectively transmitted over $T$ antennas.
The transmitted symbols are represent by a $T \times N$ matrix $\mathbf{X}$, where the rows and columns are corresponding to the transmitting antennas and time slots, respectively.
In this paper, only QAM is considered, and the average transmitting power of the transmitted symbols are normalized to one, i.e., $\mathbb{E}[\|x_{i,j}\|^2]=1$.

At the receiver, $M$ antennas are configured. Thus, the MIMO channel at the $t$-th time slot can be described by a $M \times T$ matrix $\mathbf{H}(t)$ with $t=1,2,\cdots, N$. The $t$-th column of $\mathbf{Y}$, i.e. the received signals at the $t$-th time slot is
\begin{equation}
\label{equ_channel}
\mathbf{Y}_t = \mathbf{H}(t) \cdot \mathbf{X}_t + \mathbf{Z}_t
\end{equation}
where $\mathbf{Z}$ is an $M \times N$ additive noise matrix, the elements of which are i.i.d. complex circular Gaussian random variables with mean zero and variance $\sigma ^2$, i.e. $z_{i,j} \sim \mathcal{CN}(0, \sigma^2)$.
In this paper, the channels between all the transmit/receive antenna pairs are assumed to be independent memoryless discrete-time normalized Rayleigh fast and uncorrelated fading channels, i.e., for any time slot $t$, the channel confident $h_{i,j}$ of $\mathbf{H}(t)$ satisfies $h_{i,j} \sim \mathcal{CN}(0, 1)$. We assume that an ideal channel estimation (instant values of $h_{i,j})$ and $\sigma^2$) is available at the receiver. Furthermore, due to the channel-aware property of polar coding, a precise knowledge of the noise variance $\sigma^2$ is assumed to be available at the transmitter which can be usually obtained from a feedback link.

After receiving $\mathbf{Y}$, a series of signal processes, i.e., MIMO detection, demodulation, de-interleaving and channel decoding, are used to retrieve the information bits $\hat{u}_\mathcal{A}$. These generalized ``decoding'' operations can be done in either a separately concatenated manner or a jointly combined manner.

\section{Channel Polarization Transforms}
\label{section_polarization}

In this section, after a brief review of the existing works, the channel polarization is extended to the MIMO transmission case. Under the multilevel coding framework, a three-stage channel transform is derived, which is the basis of the proposed STPCM scheme.

\subsection{Channel Polarization of $2^m$-ary PCM}

In the initial work of Ar{\i}kan \cite{Arikan}, a mapping $W: \mathcal{X} \mapsto \mathcal{Y}$ is used to denote a BMC channel, where $\mathcal{X}$ and $\mathcal{Y}$ are the input and output alphabets, respectively. Since the channel input is binary, $\mathcal{X}=\left\{0,1\right\}$. The channel transition probabilities are $W\left(y|x\right)$, $x\in \mathcal{X}$ and $y\in \mathcal{Y}$.
After channel combining and splitting operations on $N$ independent uses of $W$, we obtain $N$ successive uses of synthesized binary input channels $W^{(i)}$, $i=1,2,\cdots,N$ with transition probabilities
\begin{equation}
\label{equ_polarized_channels}
     {W}^{(i)}(y_1^N, u_1^{i-1}|u_i)=\sum\limits_{u_{i+1}^{N} \in \mathcal{X}^{N-i}}{\frac{1}{2^{N-1}}\overline{W}(y_1^N|u_1^N)}
\end{equation}
\noindent{where}
\begin{equation}
\label{equ_polarized_channels2}
    \overline{W}(y_1^N|u_1^N)=\prod\limits_{{i}=1}^{N}{W(y_{i}|x_{i})}
\end{equation}
where
\begin{equation}
\label{equ_binary_polarization}
x_{1}^{N}={\mathbf{G}} \cdot u_{1}^{N}
\end{equation}
The matrix ${\mathbf{G}}={\mathbf{B}} \cdot \mathbf{F}^{\otimes n}$, in which ${\mathbf{B}}$ is the $N \times N$ bit-reversal permutation matrix and
\begin{equation}
{\mathbf{F}}=\left[ \begin{matrix}   1 & 0  \\   1 & 1  \\ \end{matrix} \right]
\end{equation}
After this channel transform, part of the resulting channels ${W}_N^{(i)}$ with $i \in \mathcal{A}$ becomes better than the original channel $W$, i.e. ${I}({W}_N^{(i)}) > I(W)$, where function $I(\cdot)$ is the symmetric capacity (the maximum mutual information between the channel inputs and outputs under uniform input distribution); while the others with $i \in \mathcal{A}^c$ become worse, where $\mathcal{A}^c$ is the complementary set of $\mathcal{A}$.
Ar{\i}kan proved it in \cite{Arikan} that when $N$ goes to infinity, ${W}_N^{(i)} \to 1$ with $i \in \mathcal{A}$, ${W}_N^{(i)} \to 0$ with $i \in \mathcal{A}^c$, and $\frac{|\mathcal{A}|}{N} \to I(W)$.

When dealing with the polar coded modulation problem \cite{pcm}, the channel becomes $W:\mathcal{X} \mapsto \mathcal{Y}$, where $\mathcal{X}$ is the $2^m$-ary input alphabet, $\left|\mathcal{X}\right|=2^m$, and $m=1,2,\cdots$ is the modulation order.
Every $m$ bits $b_{1}^{m}\in {{\{0,1\}}^{m}}$ are modulated into a single modulation symbol $x\in \mathcal{X}$ under a specific one-to-one mapping called constellation labeling
\begin{equation}
\label{equ_constellation_labeling}
L: {{\{0,1\}}^{m}}\mapsto \mathcal{X}
\end{equation}

Thus, the channel can be equivalently written as \mbox{$W:{{\{0,1\}}^{m}} \mapsto \mathcal{Y}$} with transition probabilities
\begin{equation}
    W(y|b_{1}^{m})=W(y|L^{-1}(x))
\end{equation}
where $L^{-1}$ is the inverse mapping of $L$.
By regarding the modulation process as a special kind of channel transform, $m$ synthesized BMCs ${{W}_{j}}:\{0,1\}\mapsto \mathcal{Y}\times {{\{0,1\}}^{j-1}}$ can be obtained, where $j=1,2,\cdots ,m$, with transition probabilities
\begin{equation}
\label{equ_pcm_channel}
{{W}_{j}}\left( y,b_{1}^{j-1}\left| {{b}_{j}} \right. \right)=\sum\limits_{b_{j+1}^{m}\in {{\left\{ 0,1 \right\}}^{m-j}}}{\left( \frac{1}{{{2}^{m-1}}}\cdot W\left( y\left| b_{1}^{m} \right. \right) \right)}
\end{equation}
After that, a conventional binary-input channel polarization transform $\mathbf{G}$ is performed on each of the resulting BMCs ${{W}_{j}}$. Finally, a series of polarized BMCs $\{{W}_{j}^{(i)}\}$ is obtained, where $i=1,2,\cdots,N$ and $j=1,2,\cdots,m$.

\subsection{Three-Stage Channel Transform of STPCM}

In the $T \times M$ MIMO transmission scenario, the channel becomes $W: \mathcal{X}^T \mapsto \mathcal{Y}^M$, where $\mathcal{X}$ and $\mathcal{Y}$ are the alphabet at each transmit or receive antenna, respectively.
In order to simultaneously transmit $T$ streams, we assume the number of the receive antennas is no less than that of the transmit antennas, i.e. $M \ge T$, and the channel matrix $\mathbf{H}$ \footnote[1]{When there is no ambiguity, the time slot index $t$ of $\mathbf{H}(t)$ is omitted to ease the presentation.} is full rank.
In each time slot, after transmitting the $T$ symbols of $x_1^T$ via the $T$ transmit antennas respectively, the received signal is
\begin{equation}
y_1^M = \mathbf{H} \cdot x_1^T + z_1^M
\end{equation}
where $z_r \sim \mathcal{CN}(0, \sigma^2)$ with $r=1,2,\cdots, M$.
Suppose the data stream are detected in a successive cancellation manner, $T$ correlated channels $W_k: \mathcal{X} \mapsto \mathcal{X}^{k-1} \times \mathcal{Y}$ are obtained, where $k=1,2,\cdots,T$ and the transition function
\begin{equation}
\label{equ_mimo_polarization}
W_k(y_1^M, x_1^{k-1} | x_k) = \sum \limits_{x_{k+1}^{T} \in \mathcal{X}^{T-k}}{\left(\frac{1}{|\mathcal{X}|^{T-1}} \cdot W(y_1^M | x_1^T)\right)}
\end{equation}

\begin{figure}[!t]
  \centering
  \includegraphics[width=0.95\columnwidth]{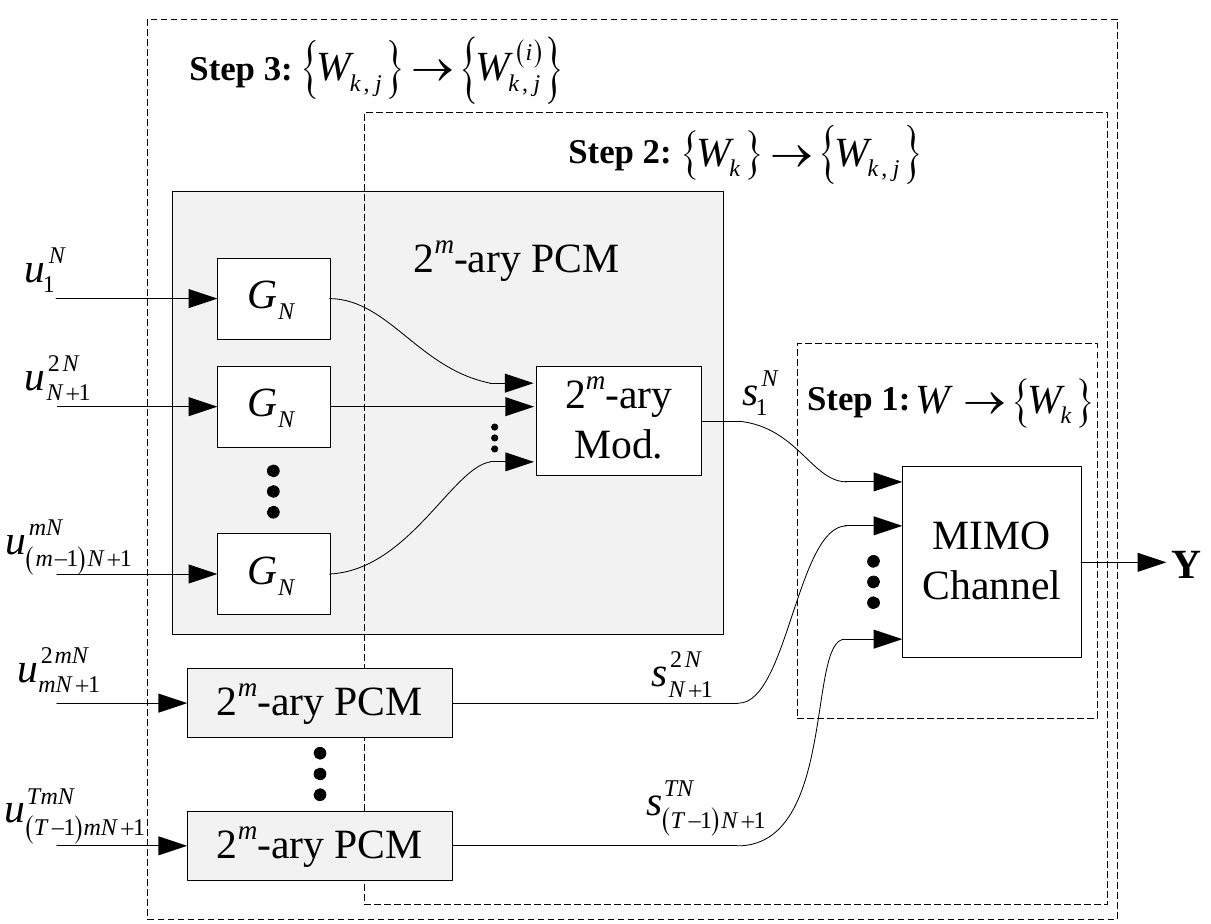}
  \caption{Three-stage channel transform of space-time polar coded modulation.}
  \label{fig_polarization}
\end{figure}

Similar to the PCM in \cite{pcm}, a three-stage channel transform is depicted in Fig.\ref{fig_polarization}.
After performing the first stage channel transform which is induced by the MIMO transmission in (\ref{equ_mimo_polarization}), each of the resulting $2^m$-ary input channels $W_k$ is transformed into a set of binary input channels $\{W_{k,j}\}$ with $j=1,2,\cdots, m$,
\begin{eqnarray}
\label{equ_wkj}
&& {{ {{W}_{k,j}}\left( y_{1}^{R},x_{1}^{k-1},b_{1}^{j-1}\left| {{b}_{j}} \right. \right)}} \nonumber \\
&& \quad =\sum\limits_{\begin{smallmatrix}
 b_{j+1}^{m}\in {{\left\{ 0,1 \right\}}^{m-j}},  \\
 {{x}_{k}}={{L}}\left( b_{1}^{m} \right)
\end{smallmatrix}}{\left( \frac{1}{{{2}^{m-1}}}\cdot {{W}_{k}}\left( y_{1}^{R},x_{1}^{k-1}\left| {{x}_{k}} \right. \right) \right)} \nonumber \\
&& \quad =\sum\limits_{\begin{smallmatrix}b_{j+1}^{m}\in {{\left\{ 0,1 \right\}}^{m-j}},\\{{x}_{k}}={{L}}\left( b_{1}^{m} \right),x_{k+1}^{T}\in {{\mathcal{X}}^{T-k}}\end{smallmatrix}}{\left( \frac{1}{{{2}^{Tm-1}}}\cdot{W\left( y_{1}^{R}\left| x_{1}^{T} \right. \right)} \right)}\qquad
\end{eqnarray}

Then, by respectively performing binary channel polarization transform on $N$ uses of $W_{k,j}$, totally $TmN$ BMCs $\{ W_{k,j}^{(i)} \}$ are obtained,
\begin{equation}
\label{equ_wkji_tmp}
\begin{aligned}
  & W_{k,j}^{\left( i \right)}\left( \mathbf{Y},u_{1}^{a-1}\left| {{u}_{a}} \right. \right) \\
 & =\sum\limits_{{{u}_{{{\mathcal{S}}}}}\in {{\left\{ 0,1 \right\}}^{N-i}}}{\frac{\prod\limits_{\begin{smallmatrix}
 i'=1,z_{1}^{T}={{\mathbf{X}}_{i'}}, \\
 b_{1}^{m}=L^{-1}\left( {{z}_{k}} \right)
\end{smallmatrix}}^{N}{{{W}_{k,j}}\left( {{\mathbf{Y}}_{i'}},z_{1}^{k-1},b_{1}^{j-1}\left| {{b}_{j}} \right. \right)}}{{{2}^{N-1}}}} \\
\end{aligned}
\end{equation}
where \mbox{$i =  1,2,\cdots,N$}, \mbox{$j = 1,2,\cdots,m$}, \mbox{$k = 1,2,\cdots,T$}, \mbox{$a=(k-1)mN+(j-1)N+i$}, and the set
\begin{equation}
{{\mathcal{S}}}=\left\{ a+1, a+2,\cdots,\left(k-1\right)mN+jN \right\}
\end{equation}
The one-to-one mapping from $u_1^{TmN}$ to $\mathbf{X}$ is jointly determined by the binary channel polarization (\ref{equ_binary_polarization}), modulation process (\ref{equ_constellation_labeling}) and the serial-to-parallel processing.

After substituting (\ref{equ_wkj}) into (\ref{equ_wkji_tmp}), a consistent representation of $W_{k,j}^{(i)}$ is obtained
\begin{equation}
\label{equ_wkji}
W_{k,j}^{\left( i \right)}\left( \mathbf{Y},u_{1}^{a-1}|{{u}_{a}} \right)=\sum\limits_{u_{a+1}^{TmN}\in {{\left\{ 0,1 \right\}}^{TmN-a}}}{\frac{\overline{W}\left( \mathbf{Y}|u_{1}^{TmN} \right)}{{{2}^{TmN-1}}}}
\end{equation}
where
\begin{equation}
\label{equ_jwkji}
\overline{W}\left( \mathbf{Y}|u_{1}^{TmN} \right) = \prod\limits_{{i}=1}^{N}{W(\mathbf{Y}_{i}|\mathbf{X}_{i})}
\end{equation}


\section{The Proposed STPCM scheme}
\label{section_stpcm}

When construct a practical STPCM scheme, similar to the conventional polar coding in \cite{Arikan} and the PCM in \cite{pcm}, the most reliable $K$ channels of $\left\{W_{k,j}^{(i)}\right\}$ are selected for carrying the information bits $u_\mathcal{A}$.
In the existing works where only AWGN channels are considered, the channel reliabilities can be evaluated efficiently by using Gaussian approximation (GA) of DE \cite{GA}.
However, the channel model considered in this paper is Rayleigh fast fading channel, and no existing practical solution is available under this scenario.

Therefore, we first propose a PCM scheme over the Rayleigh fast fading channel, where the fading channel is approximated by an AWGN channel with identical capacity.
After that, a STPCM scheme is derived based on the three-stage channel polarization discussed in the previous section.

\subsection{Channel Transform with QR decomposition}
\label{sub_ch_trans_with_qr}
In this subsection, the channel transform of the MIMO channel under a QR decomposition is proposed.
Since the channel transform in (\ref{equ_mimo_polarization}) implicates a detecting order of the data streams, the transmitter and receiver should have an agreement on the specific MIMO detection solution.

In this paper, QR-decomposition is applied to for each channel coefficient matrix $\mathbf{H}$ at the receiver,
\begin{equation}
\label{equ_qr}
\mathbf{H} = \mathbf{Q} \cdot \mathbf{R}
\end{equation}
where $\mathbf{Q}$ is an $M \times M$ unitary matrix, and $\mathbf{R}$ is an $M \times T$ upper triangular matrix with $\Im{({r_{j,j}})}=0$ for any $1 \le j \le T$ and $r_{i,j}=0$ for any $i < j$, where $1 \le i \le M$ and $T \le M$.

The received signal in (\ref{equ_channel}) after QR-decomposition detection is
\begin{equation}
\label{equ_ch_after_detec1}
\tilde{\mathbf{Y}}_t = {\mathbf{Q}^\dag} \cdot {\mathbf{Y}}_t = \mathbf{R} \cdot \mathbf{X}_t + \tilde{\mathbf{Z}}_t
\end{equation}
where the elements in $\tilde{\mathbf{Z}}_t={\mathbf{Q}^\dag} \cdot {\mathbf{Z}}_t$ is still i.i.d Gaussian distributed, $\tilde{z}_{i,t} \sim \mathcal{CN}(0,\sigma^2)$ for any $1 \le i \le M$.

After expanding the matrix operations in (\ref{equ_ch_after_detec1}), we have
\begin{equation}
\label{equ_ch_after_detec}
\tilde{y}_{k, t} = r_{k,k}\cdot x_{k,t}+\sum\limits_{k'=k+1}^{T}{r_{k,k'}\cdot x_{k',k}}+\tilde{z}_{k,t}
\end{equation}
The transmitted streams $\left\{\mathbf{X}^\prime_k\right\}$ can be detected in a decreasing order in the antenna index $k$, i.e., the stream from the $T$-th transmit antenna is first detected, then the $(T-1)$-th, $\cdots$, finally the $1$st.
Under such a successive cancellation detection, when dealing with $x_{k,t}$, the term $\sum\nolimits_{k'=k+1}^{T}\left({r_{k,k'}\cdot x_{k',k}}\right)$ in (\ref{equ_ch_after_detec}) can be dropped.
Therefore, the $\{x_{k,t}\}$ are equivalent to be transmitted over a fading channel with gains $\{r_{k,k}\}$. Thus the resulting polarized channels \{$W_k$\} in (\ref{equ_mimo_polarization}) (redefined in a reversed order of $k$) are written as
\begin{eqnarray}
\label{equ_qr_channel}
&& W_k\left(\mathbf{Y}_t, x_{k+1,t}, x_{k+2,t}, \cdots, x_{T,t} | x_{k,t}\right) \nonumber \\
&& \qquad \quad =\frac{1}{\sqrt{\pi \sigma^2}}\exp{\left(-\frac{\|\tilde{y}_{k,t}-r_{k,k}\cdot x_{k,t}\|^2}{\sigma^2}\right)}
\end{eqnarray}

At the transmitter, the detecting order of the transmit streams and the noise variance $\sigma^2$ are assumed to be notified through a feedback link.
However, the $\{r_{k,k}\}$, or equivalently $\mathbf{H}(t)$, is time-varying, and the instantaneous values of the channel coefficients are unavailable at the transmitter when the channel is assumed to be fast fading.
This is quite different from the conventional polar coding schemes where the precise knowledge of the channel state is known at both the transmitter and receiver.
The reliabilities of $\{W_k\}$ cannot be precisely evaluated by the existing solutions, so the PCM scheme in \cite{pcm} also cannot be applied directly on each $W_k$.

In the following part of this subsection, we propose to construct PCM schemes by approximating the fading channels using a set of AWGN channels which have the same capacities with the originals, and it will be used in the construction of the proposed STPCM scheme in Section \ref{subsec_multilevel_stpcm}.

The $2^m$-ary QAM with constellation $\mathcal{X}$ is equivalent to two independent $2^{m/2}$-ary pulse amplitude modulations (PAM) with constellations $\Re{(\mathcal{X})}$ and $\Im{(\mathcal{X})}$, respectively.
Without loss of generality, we assume the real and image parts of the QAM constellation are identical, i.e., $\Re{(\mathcal{X})}=\Im{(\mathcal{X})}$.

The symmetric capacity of an AWGN channel under $2^m$-ary QAM with noise variance $\sigma^2$ is
\begin{equation}
\label{equ_agwncap}
I_{G}(\sigma) = -\int_{-\infty}^{+\infty}{2p(y) \cdot \log p(y)\text{d}y} - \log{\pi e \sigma^2 }
\end{equation}
where
\begin{equation}
p(y)=\frac{1}{2^{m-1} } \sum \limits_{x \in \Re(\mathcal{X}) } { \frac{1}{\sqrt{2\pi \sigma^2}}\exp\left(\frac{-(y-x)^2}{2\sigma^2}\right) }
\end{equation}

Under the channel model described in \mbox{Section \ref{section_model}}, the channel coefficients $\{h_{i,j}\}$ are i.i.d normalized circular Gaussian distributed. According to \cite[Theorem 3.3]{QR}, twice the square of the elements $r_{k,k}$ in $\mathbf{R}$ are scaled $\chi^2$-distributed with $2(M-k+1)$ degrees of freedom, i.e., $2r^2_{k,k}\sim \chi^2\left(2(M-k+1)\right)$ for $k=1,2,\cdots, T$, where the probability density function (pdf) of $\chi^2 \left(\kappa\right)$ for a given value $\gamma$ and $\kappa$ degrees of freedom is
\begin{equation}
\label{equ_chi2}
p(\gamma, \kappa) = \frac{(1/2)^{\kappa/2}}{\Gamma(\kappa/2)}\gamma^{\frac{\kappa-2}{2}}\exp\left({-\frac{\gamma}{2}}\right)
\end{equation}
where $\Gamma(\cdot)$ is the Gamma function
\begin{equation}
\label{equ_gamma}
\Gamma \left( \kappa  \right)=\int_{0}^{+\infty }{{{e}^{-t}}{{x}^{\kappa -1}}\text{d}t}
\end{equation}
Thus, the ergodic capacity of a $T \times M$ MIMO channel $W$ under $2^m$-ary QAM with noise variance $\sigma^2$ can be calculated as
\begin{equation}
\label{equ_fadingcap}
I_{W}(\sigma) = \sum \limits_{k=1}^{T} {I_{k}{(\sigma )}}
\end{equation}
where $I_{k}(\sigma)$ is the ergodic capacity of $W_k$ which is calculated as
\begin{equation}
\label{equ_fadingcap_sub}
I_{k}(\sigma) = \int _{0}^{+\infty}{I_{G}\left({\sigma}/{a}\right)p(a)\text{d}a}
\end{equation}
with $2a^2 \sim \chi^2\left(2\left(M-k+1\right)\right)$.

After approximating the fading channel $W_k$ using an AWGN channel $\tilde{W}_{k}$ with $\sigma_{k}$, where
\begin{equation}
I_{G}(\sigma_{k}) = I_{k}(\sigma)
\end{equation}
the code construction and performance evaluation is then performed over each of the equivalent AWGN channels $\left\{\tilde{W}_{k}\right\}$ in the same way as that in the conventional AWGN case \cite{pcm}.
As that will be shown in \mbox{Section \ref{section_simulation}}, the bounds obtained by GA under this equivalence well match the simulated block error rate (BLER) curves.

\subsection{Multilevel STPCM}
\label{subsec_multilevel_stpcm}

\begin{figure*}[!t]
  \centering
  \includegraphics[width=.8\linewidth]{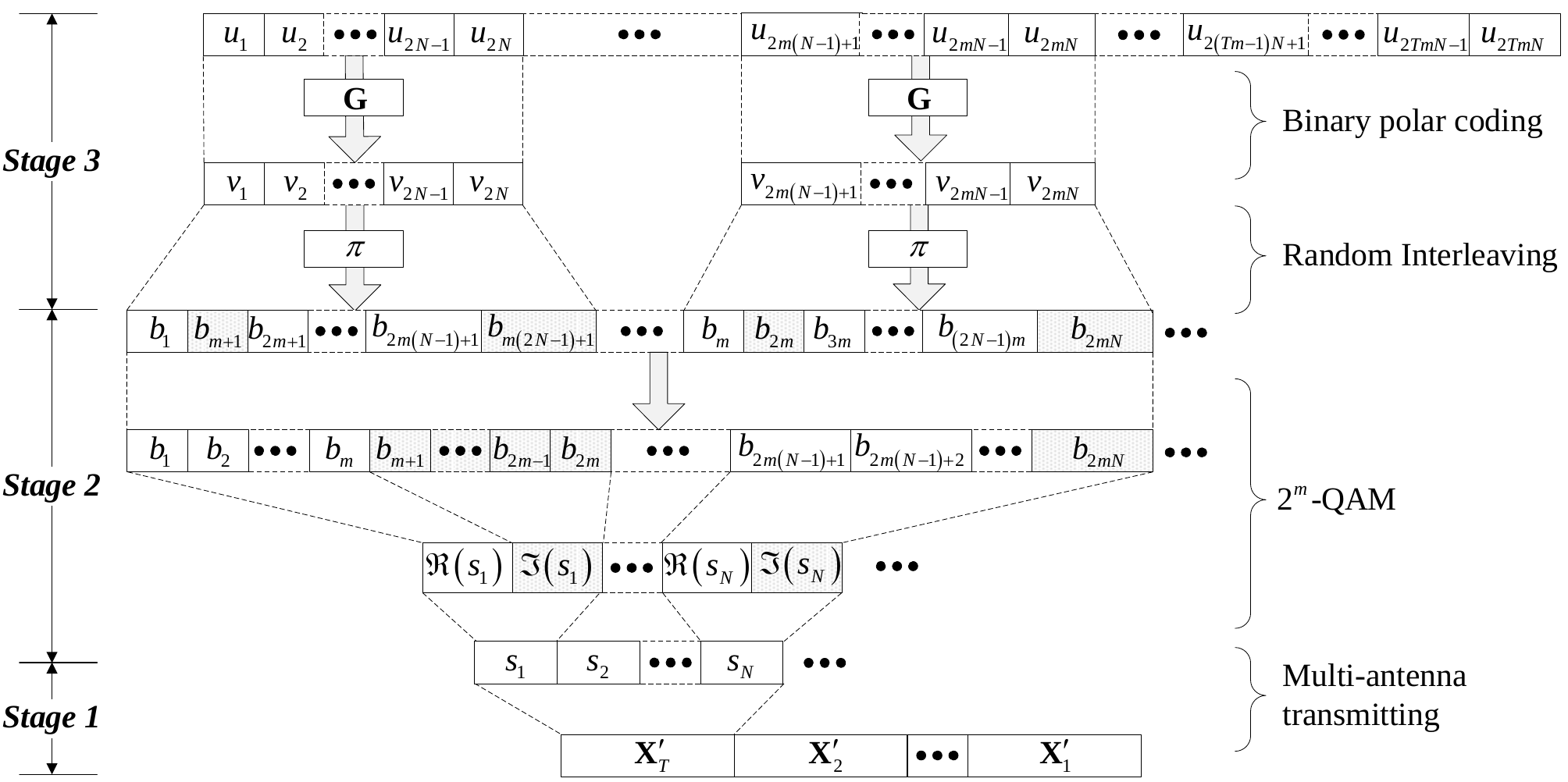}
  \caption{The mapping from $u_1^{TmN}$ (which consists of $u_{\mathcal{A}}$ and $u_{\mathcal{A}^c}$) to $\mathbf{X}$ under the proposed STPCM scheme.}
  \label{fig_frame}
\end{figure*}

In this subsection, a  STPCM scheme over QAM modulated channels base on QR-decomposition is proposed.

As for the scheme construction, $N$ independent uses of the MIMO channel $W$ in $N$ time slots are transformed into a series of binary-input channels in a three-stage channel transform.
When the code length is finite, PCM under set-partitioning (SP) labeling is find to achieve the best performance over PAM modulated channel \cite{pcm}.
In this paper, an identical SP labeling rule is applied on the real and image parts of the QAM constellation, respectively.
Moreover, to achieve the optimal utilization of the channel independencies, an additional transform is applied between the real-valued channel pairs corresponding to real and image parts of QAM symbols.

The detailed channel transform adopted by the proposed STPCM scheme is described below.

\renewcommand{\theenumi}{\arabic{enumi}}
\renewcommand{\labelenumi}{S\theenumi)}

\renewcommand{\theenumii}{\theenumi. \arabic{enumii}}
\renewcommand{\labelenumii}{(\theenumii)}

\begin{enumerate}

\item \label{S1} After a QR-decomposition of the MIMO channel $W$, a set of single-input multiple-output (SIMO) channels $\{W_k\}$ are obtained, with $k=1,2,\cdots,T$.

\item \label{S2} Under a $2^m$-ary QAM with $m=2, 4, 6, \cdots$, the SIMO channel $W_k$ at the $k$-th transmit antenna is further transformed to $m$ binary-input channels $\{W_{k,j}\}$. Without loss of generally, the first half of the channels with $j=1,2,\cdots, m/2$ correspond to the real parts of the modulation symbols, while the other half correspond to the image parts.

\item \label{S3} For each $1 \le j \le m$ and $1 \le k \le T$, $N$ uses of $W_{k,j}$ are further transformed by an $N$-scaled binary channel polarization.
    Since at the $k$-th transmit antenna, the channels corresponding to the real and image parts of the symbols, i.e. $\{W_{k,j}\}$ and $\left\{W_{k,j+m/2}\right\}$ with $j=1,2,\cdots,m/2$, are noised by independent real-valued AWGNs, and the labeling rules of the real and image parts of the QAM constellation are set to be identical, one additional step of binary channel polarization transform can be performed between the channel pairs of $\{W_{k,j}\}$ and $\{W_{k,j+m/2}\}$.
    Thus, for each $j \in \{1,2,\cdots, \frac{m}{2}\}$ and $k \in \{1,2,\cdots,T\}$, the $2N$ channel uses, which consist of $N$ uses of $W_{k,j}$ and $W_{k,j+m/2}$ respectively, are then transformed by an $2N \times 2N$ matrix $\mathbf{G}$ into $\{W_{k,j}^{(i)}\}$, where $i=1,2,\cdots, 2N$.
    Note that when the system is working over fading channels, the shared channel gain $r_{k,k}(t)$ introduces correlationship between the channel pairs $W_{k,j}$ and $W_{k,j+m/2}$ for any specific time slot $t$. Therefore, an additional interleaving process is required on the inputs of $\{W_{k,j}\}$ and $\{W_{k,j+m/2}\}$ to make the channel uses participated in the binary channel transform $\mathbf{G}$ independent.

\end{enumerate}

\medskip

Following Section \ref{sub_ch_trans_with_qr}, when constructing the STPCM scheme, each of the channels $\{W_{k}\}$ obtained in S1) are approximated using an AWGN channel, the reliabilities of the corresponding $\{W_{k,j}^{(i)}\}$ can then be evaluated by DE or GA in the same way as that in the conventional PCM scheme.
Finally, the $K$ most reliable channels among $\{W_{k,j}^{(i)}\}$ are selected for carry the information bits $u_\mathcal{A}$, and the others are fixed to frozen bits $u_{\mathcal{A}^c}$, while the universal set $\mathcal{A} \bigcup \mathcal{A}^c = \{a| \mbox{$a=(k-1)mN+2(j-1)N+i$}$, where \mbox{$i=1,2,\cdots, 2N$}, \mbox{$j=1,2,\cdots, \frac{m}{2}$}, \mbox{$k=1,2,\cdots,T$}\}.

\mbox{Fig. \ref{fig_frame}} gives an illustration of the mapping from the information bits $u_{\mathcal{A}}$ to the transmitted signal matrix $\mathbf{X}$.

At the receiver, since both the MIMO channel and the modulation procedures are combined into the channel polarization transform, the MIMO detection, demodulation and binary polar decoding can be jointly processed.
A successive cancellation (SC) algorithm can be used to decode this generalized polar code.
Given the received signal $\mathbf{Y}$, the information bit $u_a$ is decoded with indices $a$ taking values from $1$ to $TmN$ under an SC manner
\begin{equation}
\label{equ_sc}
{\hat{u}_{a}}=\left\{ \begin{matrix}
   {{h}_{a}}\left( \mathbf{Y},\hat{u}_{1}^{a-1} \right) & \text{if }a\in \mathcal{A}\text{ }  \\
   {{u}_{a}}\qquad\qquad\;\; & \text{if }a\in {{\mathcal{A}}^{c}}  \\
\end{matrix} \right.
\end{equation}
where
\begin{equation}
{{h}_{a}}\left( \mathbf{Y},\hat{u}_{1}^{a-1} \right)=\left\{ \begin{matrix}
   0 & \text{if }\frac{W_{k,j}^{\left( i \right)}\left( \mathbf{Y},\hat{u}_{1}^{a-1}|0 \right)}{W_{k,j}^{\left( i \right)}\left( \mathbf{Y},\hat{u}_{1}^{a-1}|1 \right)}\ge 1  \\
   1 & \text{otherwise} \qquad \qquad \quad  \\
\end{matrix} \right.
\end{equation}
where the indices of the channel $W_{k,j}^{(i)}$ are calculated as \mbox{$i= \left((a-1)\mod 2N\right) +1$}, \mbox{$j= \left( \left \lfloor \frac{a}{2N} \right \rfloor \mod \left(\frac{m}{2}\right) \right)+1$}, $k=  \left \lfloor \frac{a}{2mN} \right \rfloor +1$, and $\lfloor \cdot \rfloor$ is the floor function.

Equation (\ref{equ_sc}) is essentially the same with the conventional SC decoding rules in \cite{Arikan}.
Therefore, the improved SC decoding algorithms, SCL \cite{SCL:Tal}\cite{SCL} and CRC-aided SCL (CASCL) \cite{list_arxiv}\cite{acadec}\cite{CAdec} can also be used to decode the proposed STPCM scheme which can yield much better performance than SC.

\subsection{Complexity of STPCM}

The proposed STPCM scheme over $T \times M$ MIMO channel with $2^m$-ary QAM is equivalent to a set of $Tm/2$ binary polar codes with code length $2N$.

For binary polar coding, the encoding and decoding complexities are both $O\left(N \log N\right)$. Compared to the component polar encodings, the complexities brought by the modulation and interleaving operations are negligible, so the encoding complexity of STPCM is $O\left(TmN \log N\right)$.

To decode the STPCM, a QR-decomposition is applied on each $\mathbf{H}(t)$ for $t=1,2,\cdots, N$. Since it is assumed that $M \ge T$ and the complexity of QR-decomposition operation is $O\left(M^3\right)$ \cite{QRbook}, the decoding complexity of STPCM is $O\left(TmN \log N + M^3 N \right)$.

\section{Simulation Results}
\label{section_simulation}

\begin{figure}[!t]
\centering

\subfigure[The bit-interleaved turbo coded scheme.]{
   \includegraphics[width=0.95\columnwidth] {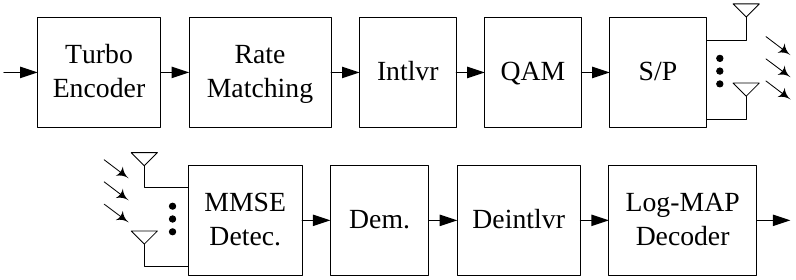}
   \label{fig_system_turbo}
 }

\subfigure[The proposed STPCM scheme.]{
   \includegraphics[width=0.95\columnwidth] {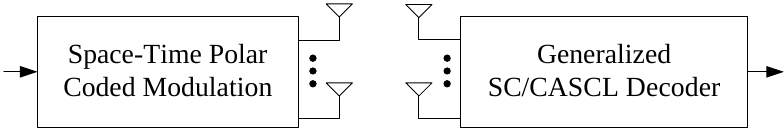}
   \label{fig_system_stpcm}
 }

\label{fig_sim_config}
\caption{The block diagrams of the simulated transmission schemes.}
\end{figure}

In this section, the BLER performance of STPCM scheme is analyzed via simulations. The number of available MIMO channel uses $N$ is in $\{128, 256, 512\}$, and the code rate \mbox{$R=\frac{K}{TmN}$} is in $\{\frac{1}{3}, \frac{1}{2}, \frac{2}{3}\}$.

For comparison, the the performance of a bit-interleaved turbo coded modulation (BITCM) scheme over \mbox{MIMO} channel is also provided. The turbo encoder and rate-matching algorithm used in 3GPP WCDMA system \cite{WCDMA} are adopted. The punctured codeword is fed into a randomized interleaver, and then modulated and distributed to the transmit antennas.
The constellation rule of BITCM scheme is Gray mapping.
At the receiver, the MMSE detection \cite{mmse}, demodulation, deinterleaving and Log-MAP decoding (with maximum $8$ iterations) \cite{logmap} are executed sequentially. This transmission model is essentially applied in many practical wireless communication systems \cite{WCDMA} \cite{LTE}.

Different from the separate signal processing in the above BITCM scheme, the propose STPCM can be regarded as a joint processing of the channel coding, modulation and the MIMO transmission. \mbox{Fig. \ref{fig_system_stpcm}} gives a block diagram of STPCM transmission. The SC decoding algorithms in (\ref{equ_sc}) is used to decode the STPCM.
As stated in \cite{CAdec}, when decoding the BITCM, $8\times2\times (8 \times 4) \times K = 512K$ metric updating operations in the trellis representations of the component convolutional codes is required: $8$ iterations over $2$ constituent $8$-state decoders with $4$ metric updates per trellis node, and the interleaver size $K=TmNR$.
When decoding the STPCM under SC, the number of required metric updates in trellis of the $\frac{Tm}{2}$ component polar codes is $TmN \log(2N)$.
Therefore, under the simulated configurations of $R$ and $N$, the BITCM consumes about $17 \sim 43$ times of the computational complexity taken by STPCM under SC decoding.

\begin{figure}[!t]
  \centering
  \includegraphics[width=.8\linewidth]{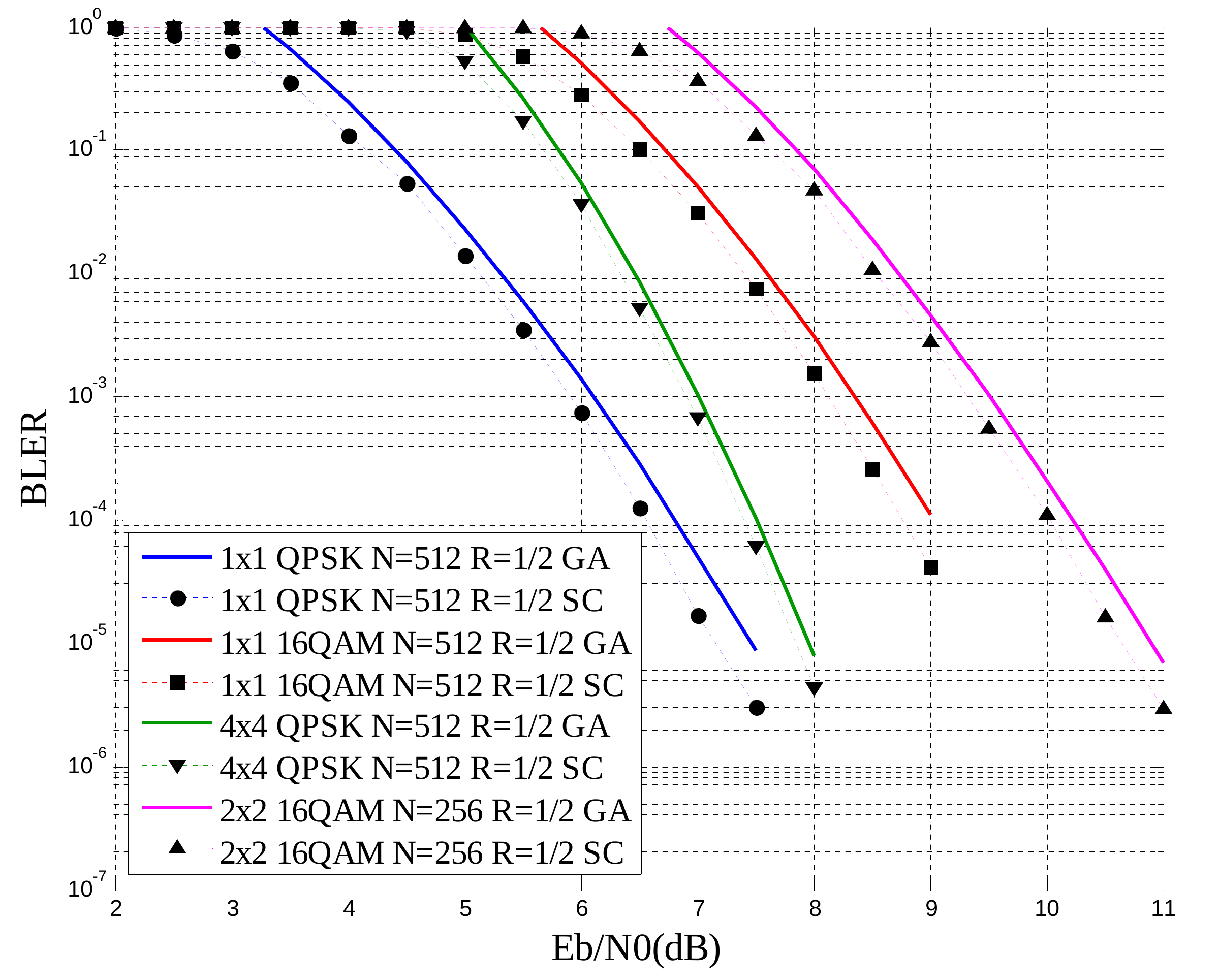}
  \caption{The simulated BLER performances over Rayleigh fast and uncorrelated MIMO channels of STPCM under SC decoding and the estimated values obtained by GA are well matched.}
  \label{fig_sim_ga}
\end{figure}

\mbox{Fig. \ref{fig_sim_ga}} gives the BLER performance of STPCM under SC decoding. The simulated STPCM schemes are constructed after evaluating the reliabilities of the polarized channels using GA algorithm.
Similar to the case of the conventional binary polar codes \cite{GA}, the BLER performances under the SC decoding of the proposed STPCM and the corresponding estimated values obtained by GA are well matched.

\begin{figure}[!t]
  \centering
  \includegraphics[width=.8\linewidth]{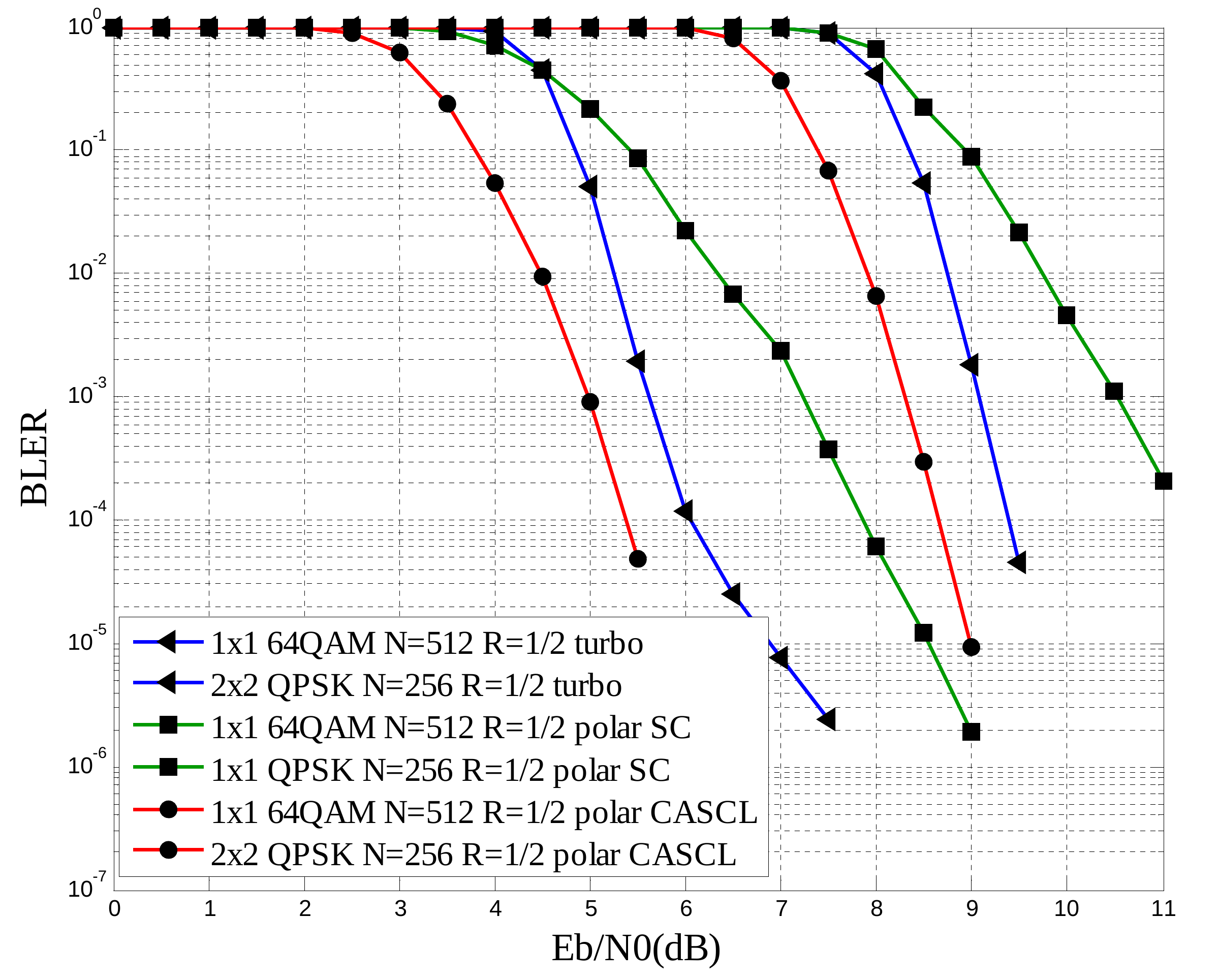}
  \caption{The performance of STPCM is significantly improved under CASCL decoding, and outperforms that of the BITCM scheme.}
  \label{fig_sim_ca}
\end{figure}

To improve the performance of STPCM, the CASCL decoding is applied.
The searching width of the CASCL decoder is set to $32$, the complexity of which is upper bounded by $32$ times of SC.
Taking the complexity-reducing implementation methods \cite{imdec_TCOM}\cite{acadec}\cite{ssc} into consideration, the CASCL decoding of STPCM under this configuration is with the comparable complexity with the Log-MAP decoding of BITCM.
As shown in \mbox{Fig. \ref{fig_sim_ca}}, the improvement in BLER performance of STPCM under the CASCL is about $2$dB or more against that under the SC decoding. Furthermore, the STPCM scheme can even outperform the BITCM scheme by no less than $0.5$dB.

\begin{figure}[!t]
  \centering
  \includegraphics[width=.8\linewidth]{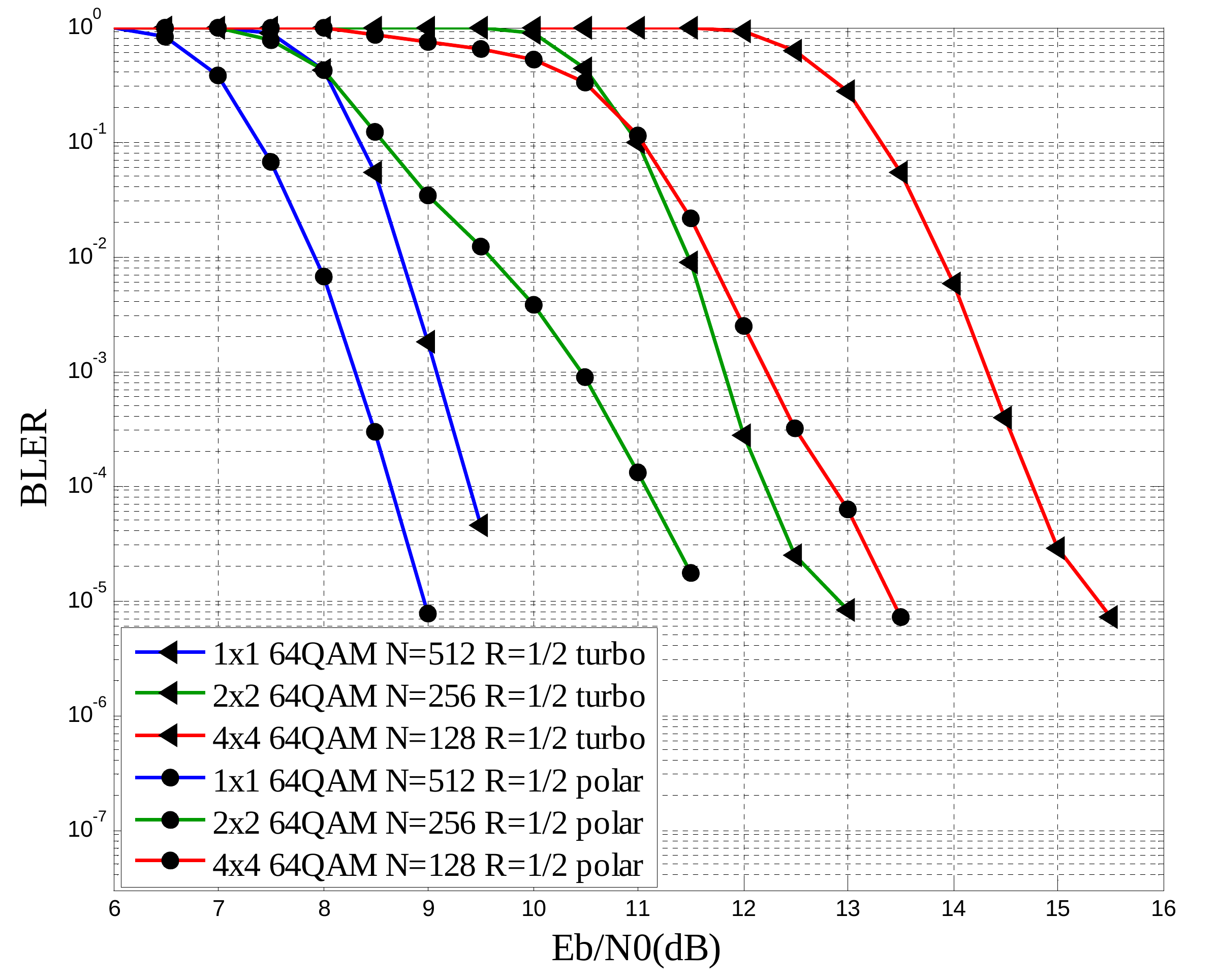}
  \caption{The simulated BLER performances over Rayleigh fast and uncorrelated MIMO channels of STPCM under different antenna settings.}
  \label{fig_sim_tr}
\end{figure}

\begin{figure}[!h]
\centering
\subfigure[ ]{
   \includegraphics[width=0.95\columnwidth] {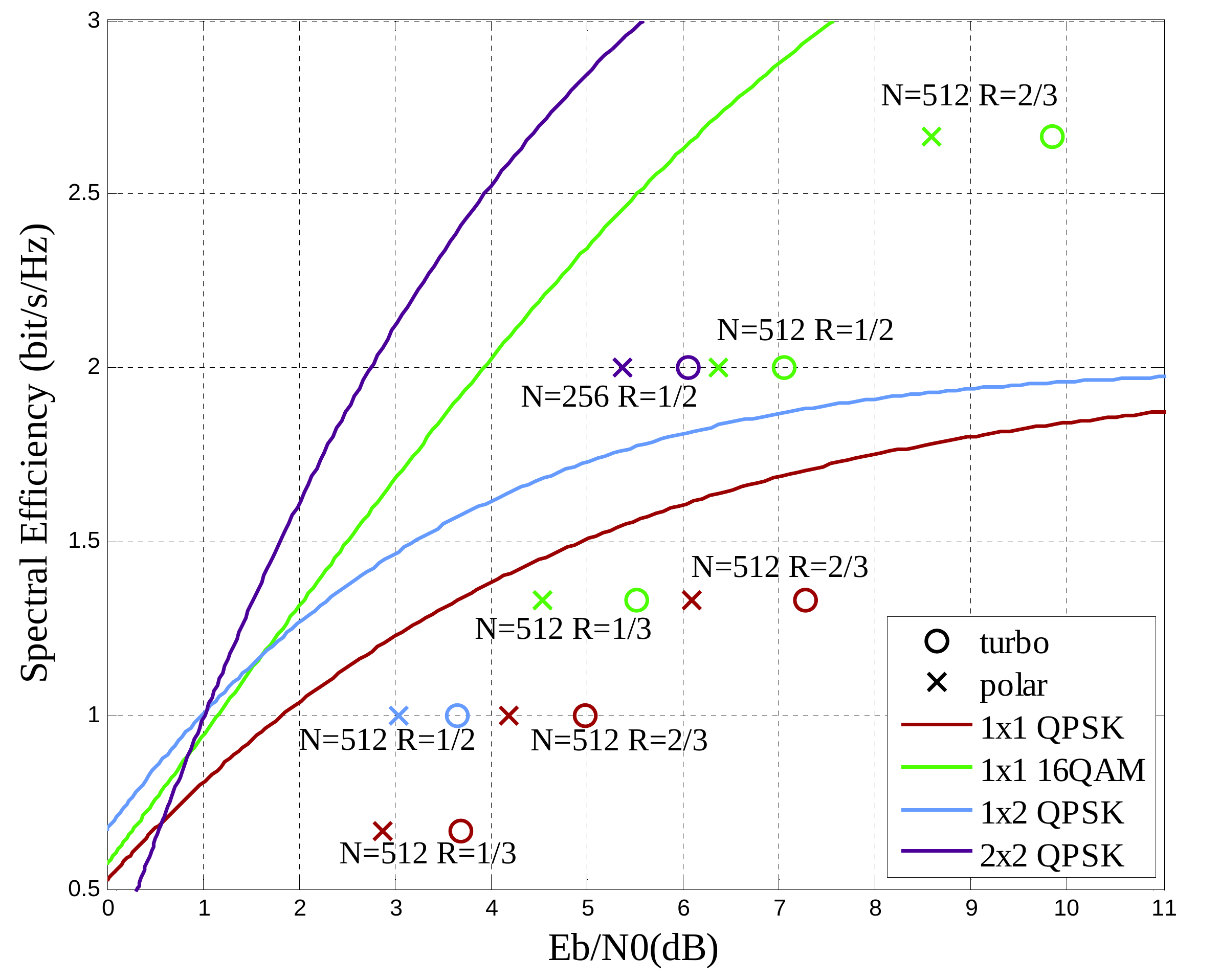}
   \label{fig_sim_cap1}
 }
\subfigure[ ]{
   \includegraphics[width=0.95\columnwidth] {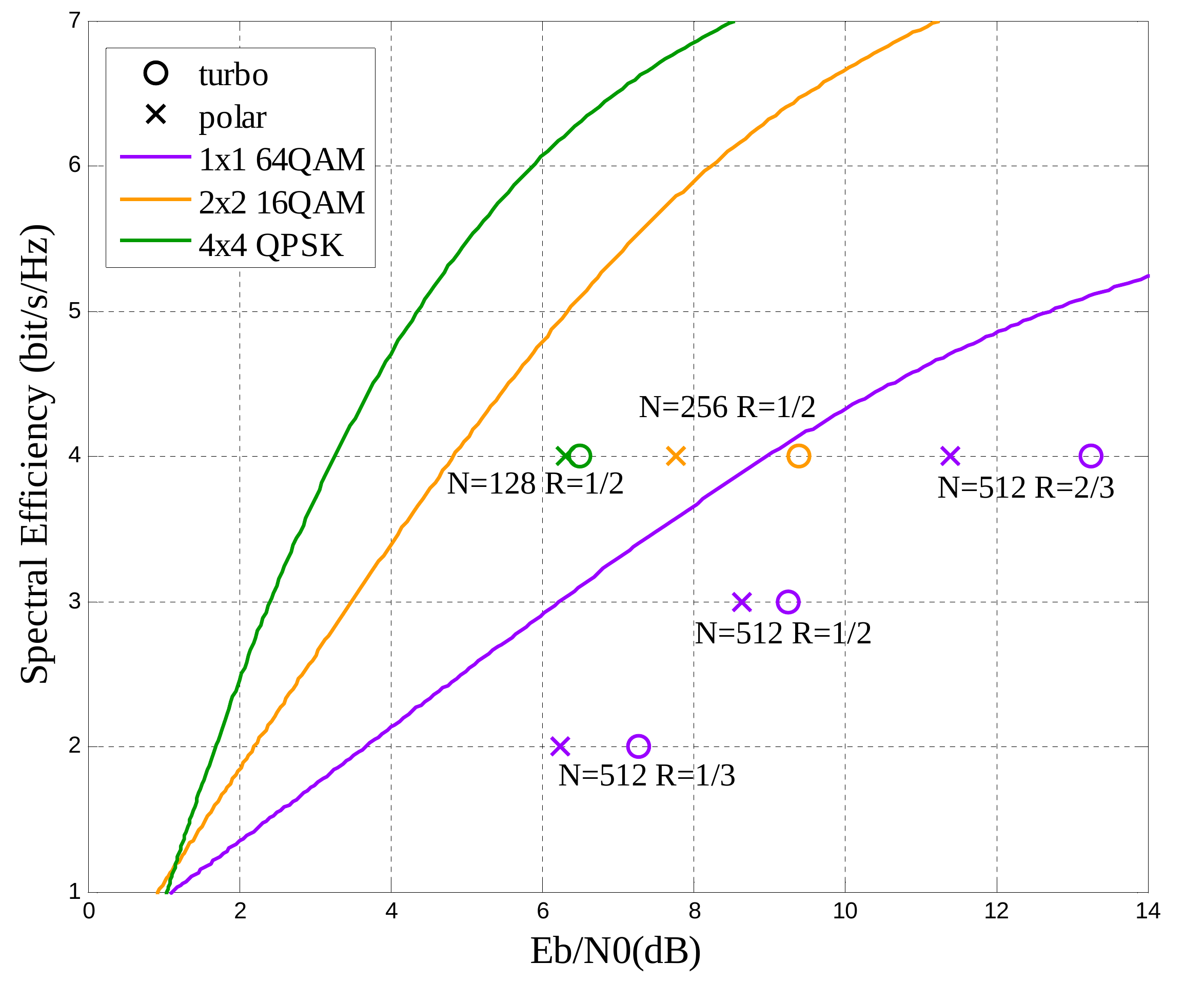}
   \label{fig_sim_cap2}
 }
\subfigure[ ]{
   \includegraphics[width=0.95\columnwidth] {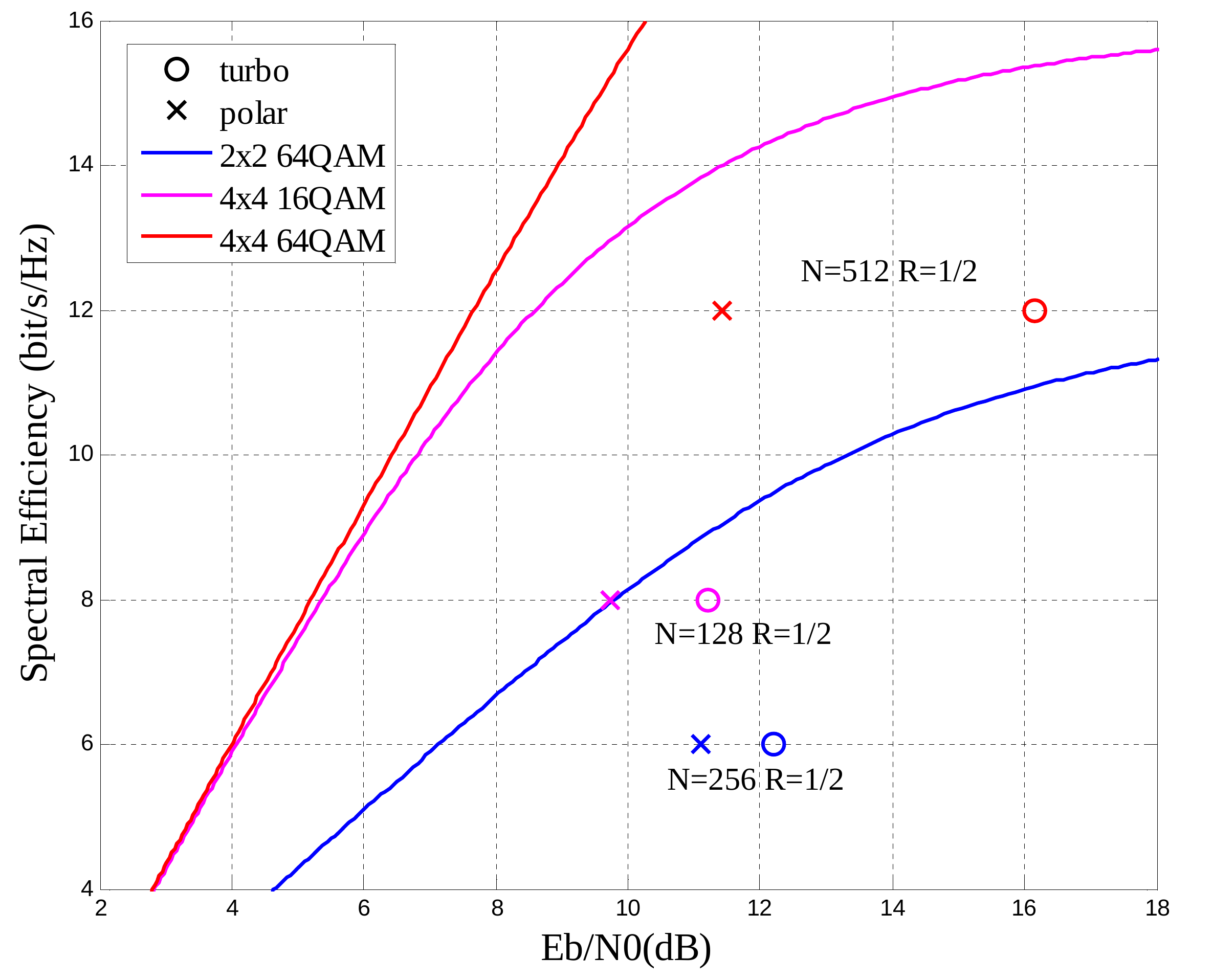}
   \label{fig_sim_cap3}
 }
\caption{The minimum required SNRs to achieve BLER $\le 10^{-4}$ over Rayleigh fast and uncorrelated MIMO channels, where the solid lines are the ergodic capacities (\ref{equ_fadingcap}) of the corresponding transmission schemes.}
\label{fig_sim_cap}
\end{figure}


The performance gain of STPCM scheme under CASCL decoding against the BITCM scheme remains when higher modulation order and more antennas are assigned. The performance over Rayleigh fast and uncorrelated MIMO channels of $64$QAM with up to $4 \times 4$ antennas are shown in \mbox{Fig. \ref{fig_sim_tr}}. As the figure shows, the performance gains are around $0.5$ to $2.0$dB.

A comprehensive comparison of STPCM and BITCM schemes under different configurations is provided in \mbox{Fig. \ref{fig_sim_cap}}. The CASCL decoding is used to decode the STPCM scheme.
In the subfigures, the minimum required SNRs to achieve BLER $\le 10^{-4}$ are plotted, and the ergodic capacities (\ref{equ_fadingcap}) of the corresponding transmission schemes are also provided. Among all the simulated cases, the STPCM scheme can achieve a performance again of $0.3$ to $2.0$dB against the BITCM scheme. Particularly, a significant $4.7$dB gain is observed in the case of $4\times4$ MIMO with $64$QAM (\mbox{Fig. \ref{fig_sim_cap3}}), when the BITCM scheme suffers with a severe \emph{error floor} effect around the BLER $10^{-4}$.

\section{Conclusion}
\label{section_conclusion}
A space-time coded modulation scheme based on polar codes is proposed following the multilevel principle, which can be seen as a joint optimization of the binary polar coding, modulation and multiple-input multiple-output (MIMO) transmission.
Similar to the multilevel approach of polar coded modulation, the MIMO transmission process is combined into the channel transform.
Based on the generalized channel polarization, a space-time polar coded modulation (STPCM) scheme with QR-decomposition is proposed for the $2^m$-ary modulated MIMO channel.
In addition, a practical solution of polar code construction over the fading channels is also provided, where the fading channels are approximated by an AWGN channel which shares the same capacity with the original.
The proposed STPCM scheme is simulated over uncorrelated MIMO Rayleigh fast fading channels.
Compared with the widely used bit-interleaved turbo coded modulation (BITCM) approach, the proposed STPCM scheme achieves a performance gain of $0.3$ to $2.0$dB in all the simulated cases.

\section*{Acknowledgment}
This work was supported in part by
the National Natural Science Foundation of China (No. 61171099),
the National Science and Technology Major Project of China (No. 2012ZX03003-007)
and Qualcomm Corporation.

\ifCLASSOPTIONcaptionsoff
  \newpage
\fi

\end{document}